# ACEM: A Cost Estimation Model for Agentic Software Engineering

Mohammad El-Ramly, Faculty of Computers and Artificial Intelligence, Cairo University.

m.elramly@fci-cu.edu.eg

https://orcid.org/0000-0002-5076-3829

Corresponding Author

## Abstract

Software cost estimation models—from COCOMO II to Function Points to Story Points—share a foundational assumption that development effort is primarily a function of human labor, primarily in designing, coding and testing software. Agentic software engineering, where autonomous AI agents perform substantial implementation work and human effort shifts from continuous production to planning, specifying, architecting, and validating agent output, challenges this assumption. New cost dimensions emerge: LLM token consumption across agent actions; Human-in-the-Loop (HITL) effort for oversight and correction; and infrastructure costs for agent orchestration and tooling. These costs operate under non-determinism—identical tasks may consume different tokens, follow divergent reasoning paths, and require varying human correction—phenomena absent in traditional development. There is a need for costing frameworks that bridge standard sizing metrics to this new cost structure.

This paper proposes ACEM (Agentic Cost Estimation Model), decomposing total agentic development cost into three additive dimensions—LLM, HITL, and infrastructure cost—without assuming a fixed hierarchy among them, since their relative magnitude varies with task complexity, agent autonomy, and pricing model. ACEM introduces three constructs for agentic dynamics: the Revision Factor (RF), modeling token overhead from output rejection and retries; the Context Factor (CF), capturing rising token consumption as context accumulates; and the HITL Intensity Score (HIS), a four-level oversight classification scheme. ACEM further maps Use Case Points, Story Points, and Function Points to estimated token consumption, letting organizations reuse existing project-scoping data for agentic cost forecasting.

ACEM is presented as a fully specified model structure and calibration methodology, with constants left symbolic pending empirical grounding; it has not yet been validated against real project data. We put ACEM forward as an early-stage proposal, together with a concrete evaluation program, and invite the community to calibrate, test, and extend it.

## Highlights

- Proposes ACEM, among the first cost estimation model designed for agentic software engineering pipelines.
- Identifies LLM token consumption as a new cost driver, beside human coding effort.
- Identifies the shift in human effort from implementation to review, oversight and correction.
- Introduces the Revision Factor (RF) and Context Factor (CF) to model retry overhead and context growth.
- Introduces the HITL Intensity Score (HIS) to classify required human oversight per task.
- Maps Use Case Points, Story Points, and Function Points to estimated token consumption.
- Presents ACEM as a theoretical framework awaiting empirical validation, with a calibration and evaluation plan proposed for future work.

## 1. Introduction

Software products are complex, invisible, malleable and adaptable. [1] This makes cost and effort estimation of software projects harder than other engineering disciplines. This has been a central concern and challenge for software engineering (SE) research and practice for decades. The ability to accurately forecast the effort, time, and resources required to deliver a software system is fundamental to project planning, resource allocation, risk management, and organizational decision-making.

Software cost estimation has a long history of model-building. From COCOMO II [2] to Function Points [3] to Use Case Points [4], the field has converged on a shared assumption: that development effort is ultimately a function of human labor, counted in person-hours or their proxies. The models differ in how they measure size and what factors they weight, but this premise runs through all of them.

That premise is now under pressure. The rise of LLM-based development tools — and more recently, autonomous agents capable of writing code, generating tests, and diagnosing bugs, proposing pull requests with minimal human direction — has begun to shift who, or what, does the work. Systems like Devin [5] and Claude Code [6] do not assist developers so much as substitute for them on defined subtasks, with the human role narrowing to oversight and correction rather than direct production. COCOMO II has no variable for this. Function points do not count tokens. The cost structure has changed, but the estimation models have not.

This transition introduces a cost structure that differs fundamentally from those addressed in traditional estimation models. In an agentic development pipeline, human coding productivity is no longer the sole cost factor traditional estimation models assumed it to be. LLM token consumption emerges as a new and, in some contexts, substantial operational cost driver — the number of input and output tokens processed by the underlying language model across all agent actions in the pipeline. Critically, the magnitude of this cost is highly variable rather than uniformly dominant: it depends on the degree of autonomy granted to the agent and the pricing model under which it operates, ranging from a negligible fraction of total project cost under light, human-supervised, subscription-based usage to a cost approaching or exceeding human labor cost under heavy, autonomous, API-billed agentic pipelines. This variability — rather than a fixed dominance of token cost over human cost — is itself a central motivation for a structured, multi-component estimation model. These token costs coexist with Human-in-the-Loop (HITL) effort [7], and with infrastructure costs required to support agent orchestration and tool integration. Critically, agentic systems also introduce non-determinism: the same task, executed twice, may consume different numbers of tokens, follow divergent reasoning paths, and demand varying levels of human correction. This variability has no precedent in classical software development and renders traditional estimation assumptions inadequate for capturing the realities of AI-driven pipelines.

Despite the rapid and wide adoption of agentic development tools in both industry and research, the software engineering community has not yet produced a systematic cost estimation framework suited to this paradigm. Existing studies on AI-assisted development, such as those evaluating the productivity impact of GitHub Copilot [8], focus on reductions in human effort rather than on the new costs introduced by LLM usage. Research on LLM cost optimization [9] addresses inference efficiency but does not connect token consumption to software engineering artifact sizing or project scope. To the best of our knowledge, no published model establishes a principled mapping between traditional software sizing metrics — such as Use Case Points or Story Points — and the token-based cost drivers of agentic development pipelines. This gap represents both a theoretical deficiency and a practical barrier: without a principled estimation framework, organizations adopting agentic SE cannot reliably forecast project costs, compare tool options, or make informed build-versus-buy decisions.

This paper is an attempt to address this gap by proposing ACEM — the Agentic Cost Estimation Model — a novel framework that extends classical software effort estimation to accommodate the cost structure of LLM-driven development. ACEM decomposes total agentic development cost into three primary dimensions: LLM token cost, HITL intervention effort, and infrastructure cost, with the first two representing the novel theoretical contributions of the model. To address the challenge of non-determinism, ACEM introduces two new corrective constructs: the Revision Factor (RF), which models the token overhead introduced by agent output rejection and retry cycles, and the Context Factor (CF), which captures the progressive increase in token consumption as context accumulates across a long agentic pipeline. To support practical adoption, ACEM further introduces the HITL Intensity Score (HIS), a four-level classification scheme for characterizing the degree of human oversight required per task, and establishes formal mappings between traditional software sizing metrics — specifically Use Case Points and Story Points — and estimated LLM token consumption, enabling practitioners to leverage existing project scoping data as input to agentic cost forecasting.

We put the model forward for the community for evaluation and validation, which would require collection of real project data from diverse settings and comparing estimations with actual costs.

The main contributions of this paper are as follows:

1. Identification and formal characterization of the novel cost drivers introduced by agentic software engineering, and demonstration of their incompatibility with existing estimation models.
2. Proposal of ACEM, a structured cost estimation framework comprising LLM cost, HITL cost, and infrastructure cost dimensions, with formal definitions of all model variables and relationships.
3. Introduction of three new estimation constructs — the Revision Factor (RF), the Context Factor (CF), and the HITL Intensity Score (HIS) — designed to address the unique characteristics of agentic development pipelines.
4. A formal mapping between traditional software sizing metrics (Use Case Points, Story Points and Function Points) and LLM token consumption, enabling integration of ACEM with existing estimation practices.

The remainder of this paper is organized as follows. Section 2 introduces the necessary background and reviews the related work spanning traditional cost estimation, agentic software engineering, and LLM cost modeling. It identifies the research gap addressed by this work. Section 3 presents the ACEM framework in full detail. Section 4 is a discussion of the model and its implications. Section 5 discusses future work and validation plans. Section 6 concludes the paper.

## 2. Background and Related work

This section reviews the body of knowledge upon which ACEM is built and situates the proposed model within the existing literature. We organize the review across four thematic areas: traditional software cost estimation (Section 2.1), AI-assisted coding (Section 2.2), agentic software engineering (Section 2.3), and LLM cost modeling (Section 2.4). Section 2.5 synthesizes these streams to identify the research gap addressed by this work.

### 2.1 Traditional Software Cost Estimation

Software cost estimation has been an active area of research since the 1970s, producing a diverse set of models that vary in their underlying assumptions, input variables, and levels of granularity. The most influential of these models can be broadly categorized into algorithmic models, sizing-based models, and expert-judgement approaches.

Algorithmic models attempt to express development effort as a mathematical function of measurable project attributes. The most widely adopted of these is the Constructive Cost Model (COCOMO), originally proposed by Boehm [10] and later extended to COCOMO II [2] to accommodate modern development practices including object-oriented design, software reuse, and iterative development. COCOMO II estimates effort in person-months as a function of software size — measured in thousands of source lines of code (KSLOC) or function points — adjusted by a set of cost drivers reflecting nature of product, platform, personnel and expertise, and project attributes. While COCOMO II remains widely used in practice, its reliance on lines of code as a primary size measure has been criticized since the inception of COCOMO as inadequate for modern development paradigms in which code generation, reuse, and abstraction obscure the relationship between size and effort [11].

Sizing-based models address this limitation by grounding estimation in functional size rather than physical code volume. Function Point Analysis (FPA), introduced by Albrecht [3], measures software size in terms of the functional transactions and data groups visible to the user, independent of implementation technology. FPA has been extensively validated across domains and remains an ISO-standardized sizing method [12]. COSMIC Function Points [13] extended the FP paradigm to better accommodate real-time and embedded systems, proposed by Karner [4] and later refined by Schneider and Winters [14], offer an alternative sizing approach grounded in use case models, making them particularly well suited to object-oriented and iterative development contexts. Use Case Points have been shown to produce effort estimates comparable in accuracy to FPA while being more accessible to practitioners working with use case-driven methodologies [15].

Agile development practices introduced a complementary estimation paradigm based on relative sizing rather than absolute measurement. Story Point estimation [16], widely adopted in Scrum-based projects, uses planning poker and team consensus to assign relative complexity weights to user stories, which are then correlated with team velocity to forecast delivery timelines. While story points lack the formal mathematical grounding of FPA, they have demonstrated practical utility in iterative, fast-moving development environments [17]. More recent work has explored hybrid approaches that combine story point estimation with machine learning models to improve accuracy [18].

Expert-judgement methods like Delphi [19] rely on expertise of team leaders and senior staff who did similar projects. To reduce bias and subjectivity, these approaches adopt structured group discussions, iterative rounds of anonymous feedback, and controlled moderation. The goal is to converge toward a consensus estimate that balances diverse perspectives while minimizing the influence of dominant voices or overconfidence. By formalizing the process, expert-judgement methods provide more reliable estimates than ad-hoc intuition, though they remain dependent on the availability of experienced professionals.

A common thread across all of these models is the assumption that effort is fundamentally a function of human cognitive work. Cost drivers in COCOMO II reflect human factors such as analyst capability, programmer experience, and team cohesion. Function points measure the functional complexity that human developers must implement. Story points reflect the collective judgment of human development teams about relative task difficulty. None of these models was designed to accommodate a development process in which a significant portion of implementation work is performed autonomously by AI agents, and in which the role of humans shift to planning, architecting, prompting, reviewing and oversight. This fundamental mismatch motivates the need for a new cost estimation framework [20].

**2.2 AI-assisted Coding**

The integration of AI into software development has progressed through several distinct phases, each characterized by a different degree of AI autonomy and a different relationship between human developers and AI systems.

Early AI-assisted development tools focused on narrow, well-defined subtasks such as code completion, syntax error detection, and automated testing. The introduction of deep learning-based code models, most notably GitHub Copilot [21] — built on the Codex model [22] — marked a significant inflection point, enabling context-aware code suggestion at the function and file level. Empirical studies evaluating the impact of Copilot and similar tools on developer productivity have reported consistent findings: developers using AI code completion tools complete tasks faster, write more code per unit time, and report higher levels of engagement [23]. Notably, Peng et al. [8] reported that developers using GitHub Copilot completed a representative programming task approximately 55% faster than a control group.

However, a critical limitation of this body of research is its framing: productivity studies in AI-assisted SE consistently measure the reduction in human effort relative to a baseline of fully human development. They do not model or measure the costs introduced by AI tool usage itself — including API costs, token consumption, prompt engineering overhead, and the cognitive effort required to evaluate and integrate AI-generated suggestions. As AI tools become more capable and more deeply integrated into development pipelines, these costs become increasingly significant and cannot be treated as negligible overheads. Additionally, these studies do not account for the long term maintenance cost of AI-generated code, relative to human-generated code.

Research on the quality of LLM-generated code has also grown substantially in recent years. Studies using benchmarks such as HumanEval [22], MBPP [24], and SWE-bench [25] have evaluated the ability of LLMs to solve programming problems of varying complexity. While state-of-the-art models demonstrate strong performance on isolated coding tasks, their accuracy degrades on tasks requiring multi-file reasoning, long-range dependency tracking, and iterative debugging [26]. These findings are directly relevant to cost estimation: lower agent accuracy implies higher revision rates, more retry cycles, and consequently greater token consumption and HITL effort — relationships that ACEM formalizes through the Revision Factor construct.

**2.3 Agentic Software Engineering**

The concept of software development agents — autonomous systems capable of planning, executing, and reflecting on multi-step development tasks — represents a qualitative leap beyond AI-assisted tools. Where AI-assisted tools augment human developers by suggesting or completing discrete actions, in agentic software engineering [27, 28], agentic systems operate with a degree of autonomy that allows them to analyze given intent, decompose and plan complex tasks, invoke external tools, maintain working memory across steps, revise their own outputs in response to feedback and collaborate with human teammates [29].

The theoretical foundations of agentic AI systems draw on research in autonomous agents [30], planning under uncertainty [31], and tool-augmented language models [32]. Practically, agentic SE systems are implemented through frameworks such as LangGraph [33], AutoGen [34], and CrewAI [35], which provide scaffolding for multi-agent coordination, tool integration, and state management. These frameworks enable the construction of development pipelines in which specialized agents collaborate on distinct aspects of the development lifecycle — for example, a requirements agent that elaborates user stories, a coding agent that implements them, a testing agent that generates and executes test cases, and a review agent that assesses code quality.

Prominent agentic SE systems that have attracted research attention include SWE-agent [36], which demonstrated the ability of LLM-based agents to autonomously resolve real GitHub issues; Devin [37], a commercially deployed agentic developer capable of end-to-end task completion across planning, coding, and deployment; and Claude Code [38], a terminal-based agentic coding assistant that operates directly within developer workflows. Evaluations of these systems on standardized benchmarks such as SWE-bench [25] have shown rapid capability improvements, with state-of-the-art agents resolving a substantial and growing fraction of real-world software issues autonomously.

A defining characteristic of agentic SE that distinguishes it from both traditional development and AI-assisted development is the role of the human developer as an intermittent supervisor rather than a continuous producer. Human-in-the-loop (HITL) interaction in agentic pipelines typically occurs at defined checkpoints — for example, after requirements elaboration, after initial code generation, or after test execution — rather than continuously throughout the development process [39]. The frequency, nature, and effort of these interactions vary significantly across task types, agent capabilities, domain risk levels, and organizational policies, and represent a major source of cost variability that existing estimation models are entirely unequipped to address.

Non-determinism is another defining characteristic of agentic SE with direct implications for cost estimation. Unlike deterministic software systems, LLM-based agents may produce substantially different outputs, follow different reasoning paths, and consume different numbers of tokens when presented with identical inputs across different runs [40]. This variability arises from the stochastic nature of LLM sampling, sensitivity to prompt phrasing, and the compounding of small differences across multi-step agent pipelines. The practical consequence for cost estimation is that point estimates of token consumption are inherently uncertain, and any viable estimation model must incorporate mechanisms for representing and managing this uncertainty — a requirement addressed in ACEM through the Revision Factor and Context Factor constructs.

### 2.4 LLM Cost Modeling

The operational cost of deploying LLM-based systems is determined primarily by token consumption, where tokens represent the fundamental unit of LLM input and output processing. Commercial LLM providers including OpenAI, Anthropic, and Google price their APIs on a per-token basis, with separate rates for input and output tokens and significant price variation across model tiers [41]. Pricing also varies by roughly two orders of magnitude across model tiers from the same provider, and providers apply additional mechanisms such as prompt caching discounts and batch-processing rates that further complicate cost prediction for a given workload. Lumer et al. [42] evaluated prompt caching strategies across OpenAI, Anthropic, and Google, showing 41–80% API cost reductions and 13–31% latency improvements for agentic workloads.

For agentic systems that execute long, multi-step pipelines involving repeated model invocations, tool calls, and context accumulation, token costs can scale rapidly and unpredictably, making cost forecasting a practically important challenge. Empirical evidence supports this concern directly: agentic coding tasks consume roughly three orders of magnitude more tokens than single-turn code reasoning or chat interactions, with input tokens — driven by accumulated context rather than generated output — accounting for the majority of cost. Critically, this consumption is not merely high but volatile: repeated executions of the identical task by the same model can differ by up to 30x in total tokens consumed, and higher token expenditure does not reliably correlate with higher task success [43]. This volatility is compounded in multi-agent configurations, where coordination overhead across agents can multiply token spend relative to single-agent pipelines without a proportional gain in output quality.

The magnitude of LLM cost relative to human labor cost in agentic development is a live and unsettled empirical question, and recent industry data suggests it varies by an order of magnitude or more depending on usage pattern. At the light end, enterprise telemetry reported by Anthropic indicates that typical Claude Code usage costs approximately $13 per developer per active day, or $150–250 per developer per month, with the large majority of users spending under $30 on any given active day — a small fraction of a typical developer's fully loaded labor cost [44]. At the heavy end, Gartner's 2026 enterprise forecast reports that nearly one-quarter of surveyed technology leaders are already spending $200–500 per developer per month on AI coding tokens, with roughly 6% exceeding $2,000 per developer per month, and projects that AI coding token costs will overtake average developer salaries by 2028 as autonomous, consumption-billed agentic workflows displace chat-assisted, subscription-billed usage [45]. The mechanism behind this divergence is directly relevant to ACEM's design: in low-autonomy, chat-assisted usage, a human remains in continuous control and token consumption per task is small and bounded, whereas in high-autonomy agentic pipelines a single task can trigger five to thirty separate model invocations, each resending accumulated conversation history and codebase context, so that cost compounds with pipeline length in exactly the manner the Context Factor (Section 3.2.4) is designed to capture. Reported cases of agentic token spend reaching tens of thousands of dollars in a single week of ungoverned autonomous use [46] illustrate the tail risk this compounding can produce. This body of evidence supports two conclusions relevant to ACEM's structure: first, that LLM token cost cannot be assumed negligible in agentic SE, as much of the AI-assisted productivity literature reviewed in Section 2.2 implicitly does; and second, that it also cannot be assumed dominant over human cost as a general rule, since its magnitude is highly sensitive to autonomy level, pipeline design, and pricing model. This variability is precisely what motivates treating costs $C_{LLM}$ and $C_{HITL}$ as independent, additive components (Section 3.1) rather than assuming a fixed cost hierarchy between them.

Research on LLM cost optimization has explored several technical strategies for reducing token consumption without proportional reductions in output quality. These include prompt compression techniques that reduce input token counts while preserving semantic content [47], KV cache sharing across concurrent requests to reduce redundant memory and recomputation overhead [48], speculative decoding approaches that use a smaller draft model to accelerate inference without altering output quality [49], and model routing strategies that dynamically assign tasks to cheaper models when full model capability is not required [50]. While these techniques are valuable for reducing per-token costs, they address the operational efficiency of deployed systems rather than the upfront estimation of costs for planned development projects — the problem domain of ACEM.

A smaller but growing body of work has examined the relationship between task complexity and LLM resource consumption in software engineering contexts specifically. Contrary to the assumption that token consumption scales with task complexity in a straightforwardly predictable manner, recent empirical evidence suggests the opposite: token usage in agentic coding tasks is highly variable and inherently stochastic, with weak correlation between conventional notions of task difficulty and actual resource consumption [43]. Complementary evidence from controlled studies of LLM-assisted development further shows that traditional effort proxies such as Story Points explain only a partial share of variance in observed effort, with validation and correction activities — rather than task size — emerging as dominant cost drivers [20]. This unpredictability has not been formalized in the context of software engineering artifact sizing, and no published work has established empirical mappings between standard SE sizing metrics and token consumption distributions for representative development tasks. This constitutes the most direct and specific antecedent gap for the contribution of ACEM

**2.5 Research Gap**

The review presented in the preceding sections reveals a clear and consequential gap in the existing literature. Traditional software cost estimation models — including COCOMO II, Function Point Analysis, Use Case Points, and Story Point estimation — provide mature, validated frameworks for forecasting development effort in human-driven projects, but are structurally incapable of modeling the cost dynamics of agentic development pipelines. AI-assisted SE research has documented the productivity benefits of LLM-based coding tools but has not addressed the costs introduced by these tools, treating token consumption and prompt engineering overhead as outside the scope of effort estimation. Agentic SE research has characterized the capabilities and limitations of autonomous development agents but has not produced cost models or estimation frameworks. LLM cost optimization research has addressed the efficiency of deployed systems but has not connected token economics to software engineering artifact sizing or project scope.

The result is that organizations adopting agentic SE tools currently have no principled basis for estimating the cost of agentic development projects. They cannot translate a project specification — expressed in use cases, user stories, or function points — into a forecast of LLM token consumption or HITL effort. They cannot quantify the cost impact of agent non-determinism or context accumulation. They cannot classify tasks by their human oversight requirements or estimate the rework costs associated with agent output rejection. And they cannot integrate agentic cost forecasting with their existing estimation practices, which are built around traditional sizing metrics.

This wide gap calls for dedicated cost estimation models for agentic software engineering. Alaswad et al. [20] proposed a conceptual framework arguing that LLM-assisted development invalidates the assumptions underlying COCOMO, Function Points, and Story Points, identifying interaction management, validation, and correction as emerging cost dimensions. In a follow-up empirical study, Alaswad et al. [51] tested this claim directly: across 22 developers completing 110 real-world tasks on three LLMs (GPT-4o, Gemini 2.5 Flash, DeepSeek-R1), they compared Story Points against a new construct, Hybrid Intelligence Effort (HIE) — a composite of five interaction-level dimensions including reasoning complexity, context completeness, iterative reasoning cycles, and human oversight effort. Story Points alone explained 72% of effort variance; adding HIE raised this to 80%, with human validation and manual correction — not raw code volume or task size — emerging as the dominant cost driver. Their central finding is that effort in LLM-assisted development shifts from construction to verification, a shift that size-based estimation models fail to capture.

These two papers represent the closest existing work to ours, but they leave a specific, self-acknowledged gap open. The conceptual framework [20] diagnoses the problem without proposing a concrete estimation model. The empirical follow-up [51] goes further by quantifying LLM-assisted effort — but does so in developer time and interaction-log counts, not token consumption, and the authors explicitly flag token-based extensions as future work outside their study's scope.

ACEM addresses precisely this gap. Rather than modeling effort in time, it establishes a direct, quantitative mapping between standard SE sizing metrics — Function Points, Use Case Points and Story Points — and token-based cost drivers, making it, to the best of our knowledge, among the first frameworks to connect artifact-level sizing to the actual operational cost unit of agentic pipelines. In doing so, ACEM introduces estimation constructs specifically designed to capture the non-determinism, context accumulation, and human-oversight dynamics that characterize agentic SE workflows. We put ACEM forward to the community for empirical evaluation and validation.

## 3. The ACEM Framework

This section presents the Agentic Cost Estimation Model (ACEM) in full detail. Section 3.1 provides an overview of the model's architecture and scope. Section 3.2 defines the LLM cost component and its constituent constructs. Section 3.3 defines the HITL cost component. Section 3.4 presents the infrastructure cost component. Section 3.5 establishes the mapping between traditional software sizing metrics and ACEM input variables. Section 3.6 describes the model calibration process. Section 3.7 gives detailed illustrative examples of ACEM application. Section 3.8 discusses model assumptions and scope boundaries.

### 3.1 Model Overview

ACEM models the total cost of an agentic software development project as the sum of three distinct cost dimensions:

$$Total_Cost = C_{LLM} + C_{HITL} + C_{Infra} \qquad (1)$$

Where:

- $C_{LLM}$ is the cumulative cost of LLM token consumption across all agent actions in the development pipeline
- $C_{HITL}$ is the cumulative cost of human-in-the-loop intervention, including review, approval, correction, and rework activities
- $C_{Infra}$ is the cost of the computational infrastructure supporting agent orchestration, tool execution, and environment management

The three dimensions are treated as additive components, reflecting distinct streams of cost (LLM tokens, HITL oversight, and infrastructure). This additivity is a pragmatic simplification rather than a claim of strict independence or orthogonality, since rejection events and context growth often affect multiple components simultaneously. The assumption holds most clearly when infrastructure is provisioned on a pay-per-use basis — as is typical in cloud-hosted agentic pipelines — and when HITL activities do not directly trigger additional LLM invocations beyond those already captured in the Revision Factor (defined in Section 3.2.3). A single rejection event typically produces two distinct, non-substitutable costs rather than one: the token cost of the agent's retry, captured in $C_{LLM}$ through the Revision Factor, and the human time to detect and direct that retry, captured in $C_{HITL}$ through $C_{rework}$ (Section 3.3.1). These are not alternative ways of accounting for the same event — both are real and are expected to co-occur — so no double-counting arises between $C_{LLM}$ and $C_{HITL}$ from a shared rejection rate $r_i$ ; rather, $r_i$ drives two genuinely separate resource streams simultaneously. This independence also gives flexibility in applying the model. For simple low autonomy tasks when LLM cost is ~1-2%, $C_{LLM}$ can be ignored; but for heavy highly autonomous pipelines $C_{LLM}$ is expected to match $C_{HITL}$ [45], both components are calculated.

Figure 1 illustrates the overall structure of ACEM, showing the relationships between model inputs, cost components, corrective factors, and the total cost output.

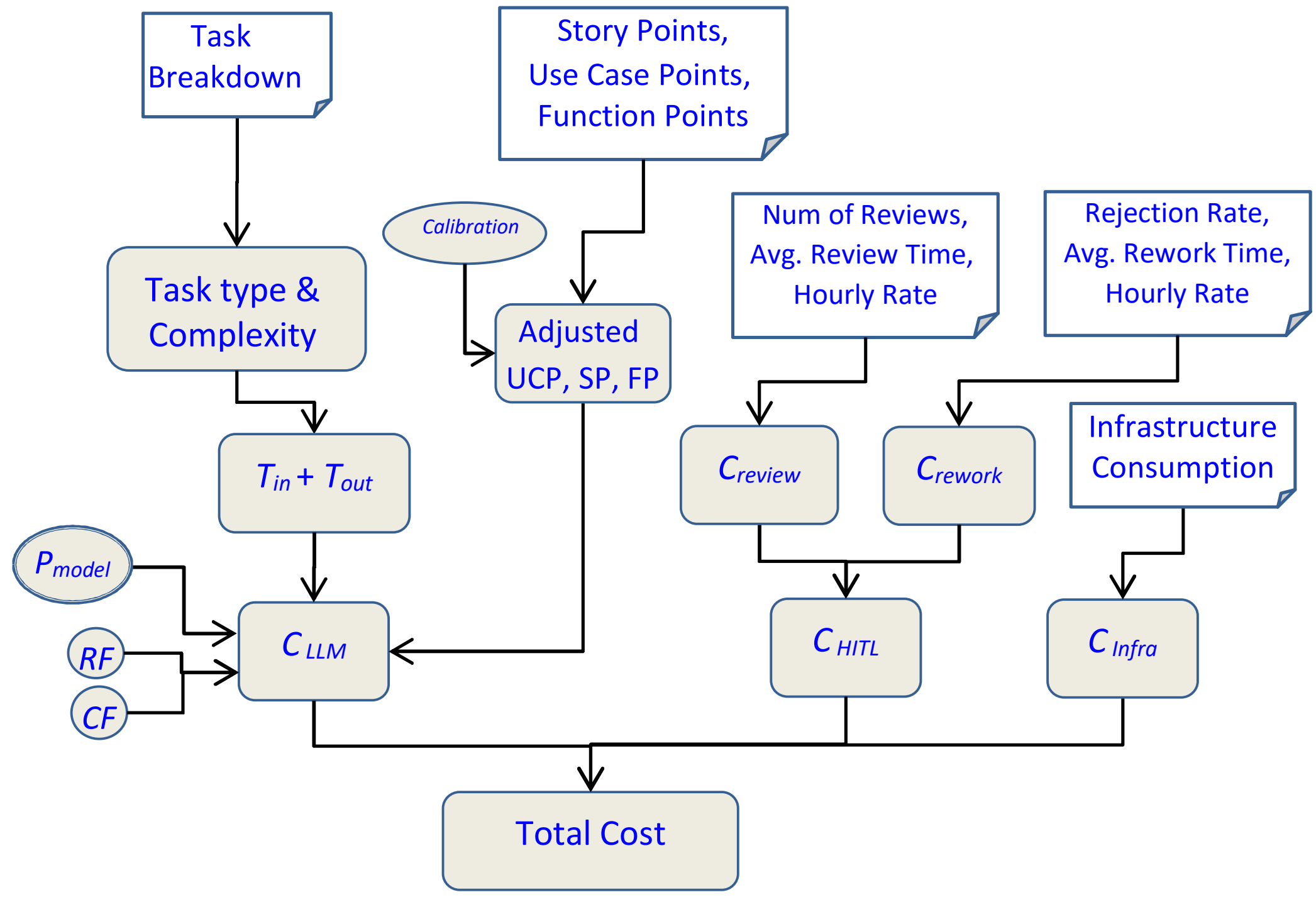

**Figure 1. Overview of ACEM Cost Estimation Model for Agentic Software Engineering**

The primary theoretical contributions of ACEM lie in the $C_{LLM}$ and $C_{HITL}$ components, which introduce novel constructs not present in existing estimation model. The $C_{Infra}$ component, while included for completeness, follows established cloud cost modeling approaches and is not a novel contribution of this work. The remainder of this section focuses accordingly on $C_{LLM}$ and $C_{HITL}$.

### 3.2 LLM Cost Component

The LLM cost component captures the total cost of token consumption across all agent actions performed during the development of a software project. An agent action is defined as any invocation of an LLM by an agent in the pipeline, including planning steps, code generation steps, test generation steps, review steps, tool call formulation steps, and reflection or self-correction steps. ACEM operates across four levels of granularity: a project decomposes into tasks (e.g., a user story or use case), each task is executed through one or more agent actions (individual LLM invocations, as defined above), and agent actions may in turn invoke tool calls. $C_{LLM}$ and $C_{HITL}$ are computed at the agent-action and task level respectively and then summed to the project level; where a sizing metric (Section 3.5) is used instead, this hierarchy is collapsed into a single project-level estimate via the calibrated γ constants.

#### 3.2.1 Base Formula

The LLM cost for a project consisting of $N$ agent tasks is defined as:

$$C_{LLM} = \Sigma_i{}^N{}_{=1} \, [(T_{in,\,i} + T_{out,\,i}) \times RF_i \times CF_i] \times P_{model} \qquad (2)$$

Where:

- $T_{in,\,i}$ is the estimated number of input tokens for task $i$
- $T_{out,\,i}$ is the estimated number of output tokens for task $i$

- $RF_i$ is the Revision Factor for task $i$, defined in Section 3.2.3
- $CF_i$ is the Context Factor for task $i$, defined in Section 3.2.4
- $P_{model}$ is the price per token for the LLM model used after caching and other reductions
- $N$ is the total number of agent tasks in the development pipeline

However in practice, a pipeline might use multiple models — for example, a more capable and expensive model for architecture planning and a lighter model for test generation — Equation 2 is extended to sum across model tiers as in Equation 3.

$$C_{LLM} = \Sigma_j \Sigma_{i \in S_j} [(T_{in,i} + T_{out,i}) \times RF_i \times CF_i] \times P_j \quad (3)$$

Where

- $S_j$ is the set of tasks assigned to model tier $j$
- $P_j$ is the per-token price of that tier after caching and other reductions

To account for the typical cases when the price of input tokens is different than the price of output tokens, Equation 3 can be adjusted as in Equation 4.

$$C_{LLM} = \Sigma_j \Sigma_{i \in S_j} (T_{in,i} \times P_{in,j} + T_{out,i} \times P_{out,j}) \times RF_i \times CF_i \quad (4)$$

- $P_{in,j}$ is the price per input token for model tier $j$ after caching and other reductions
- $P_{out,j}$ is the price per output token for model tier $j$

**3.2.2 Base Token Estimation**

The base token consumption for a task $i$ is estimated as a function of the artifact type being produced and its complexity level:

$$T_{in,i} + T_{out,i} = \text{BaseTokens}(type_i, complexity_i) \quad (5)$$

In practice, BaseTokens yields a pair of values — input and output tokens — recorded and calibrated separately, since Equation 4 prices them differently. Table 2 lists a single β symbol per cell for notational simplicity; each β expands to a calibrated ($\beta_{in}$, $\beta_{out}$) pair once populated

$T_{in,i}$ as estimated by BaseTokens is task-intrinsic: it represents the tokens specific to task $i$'s own content — its instructions, target artifact, and immediately relevant context — and explicitly excludes the accumulated conversation history and prior-step outputs carried forward from earlier tasks in the pipeline. This exclusion is deliberate: context accumulation is modeled separately and multiplicatively through the Context Factor (Section 3.2.4), so that BaseTokens can be calibrated once per artifact-type / complexity cell without being contaminated by where in the pipeline a task happens to fall

Artifact type captures the nature of the development output — for example, a user story elaboration, a function implementation, a unit test suite, or a code review. Complexity level reflects the inherent difficulty of the task, classified using a three-level scale: Simple, Medium, and Complex, defined in Table 1. Table 2 shows the base tokens per task type and complexity level.

Complexity, as used in ACEM, is not a single unambiguous quantity. An industry analysis [52] identifies the factors that shape an LLM's effective processing burden — and hence token consumption — for a given task, showing that this burden is determined by multiple interacting factors: the length of user input and model output (with output often priced higher), the cumulative size of conversational context (captured separately in ACEM's Context Factor), and the inherent difficulty of the task itself (e.g., long-form text generation, multi-step reasoning, or multimodal agent workflows). Additional influences include formatting and special characters, language and encoding differences, and model-specific behaviors such as verbosity or internal optimization level. Advanced model features further complicate this picture: extended "deep thinking" reasoning chains and online retrieval increase token consumption, while effective prompt caching reduces it. Together, these findings show that token consumption is not simply a function of human-perceived task complexity — the dimension traditional software engineering practice normally means by "complexity," and the dimension Table 1 was originally built to capture — but instead emerges from a distinct mix of input/output length, context growth, task type, language and formatting, model architecture, and optional model features. Call this second, model-facing dimension LLM-perceived complexity. Complementary evidence from code-specific analysis further questions human-centric complexity metrics: Xie et al. [53] show that, after controlling for code length, classical complexity metrics (e.g., cyclomatic complexity) exhibit no consistent correlation with LLM task performance, proposing an entropy-based alternative (LM-CC) that correlates more strongly. While their focus is model performance rather than token consumption directly, this reinforces the broader theme that human-perceived complexity is an unreliable proxy for LLM-facing difficulty.

This directly addresses a tension noted in Section 2.4: Bai et al. [43] report weak correlation between conventional task difficulty and actual token consumption. ACEM does not dispute this. Table 1 separates two dimensions — human-perceived complexity, assessable during planning via familiar SE criteria, and LLM-perceived complexity, which is agent- and model-specific. Human-perceived complexity organizes the BaseTokens table (Table 2), but each cell's actual token relationship is established empirically through calibration (Section 3.6), not assumed from the label. Where calibration shows poor predictive power for a given agent or domain, this should surface as wider confidence intervals on Table 2's β values, a shift to the reduced calibration scheme described below, or greater reliance on *RF* and *CF* to absorb residual variability. This is an explicit scope boundary: Table 2's predictive validity is an empirical question calibration is designed to answer, not an assumption the model relies on.

**Table 1: Complexity Classification Criteria**

| Level | Human-Perceived Complexity | LLM-Perceived Complexity |
|---|---|---|
| Simple | Single responsibility, no external dependencies, well-defined input/output, no ambiguity in requirements | To be calibrated empirically (Section 3.6); expected to correlate with short input/output length, minimal context, and low reasoning depth |
| Medium | Multiple responsibilities or moderate dependencies, some ambiguity requiring agent inference, moderate output length | To be calibrated empirically (Section 3.6); expected to correlate with moderate context accumulation and occasional extended reasoning |
| Complex | Cross-cutting concerns, multiple external dependencies, significant ambiguity, iterative reasoning required, long output | To be calibrated empirically (Section 3.6); expected to correlate with long output, deep reasoning chains, and/or retrieval calls, but not assumed a priori |

**Table 2: BaseTokens Reference Table**

| Artifact Type | Simple | Medium | Complex |
|---|---|---|---|
| Requirements elaboration (per user story) | $\beta_{11}$ | $\beta_{12}$ | $\beta_{13}$ |
| Use case specification (per use case) | $\beta_{21}$ | $\beta_{22}$ | $\beta_{23}$ |
| Function / method implementation | $\beta_{31}$ | $\beta_{32}$ | $\beta_{33}$ |
| Class / module implementation | $\beta_{41}$ | $\beta_{42}$ | $\beta_{43}$ |
| Unit test suite (per function) | $\beta_{51}$ | $\beta_{52}$ | $\beta_{53}$ |
| Integration test scenario | $\beta_{61}$ | $\beta_{62}$ | $\beta_{63}$ |
| Code review (per module) | $\beta_{71}$ | $\beta_{72}$ | $\beta_{73}$ |
| Bug diagnosis and fix | $\beta_{81}$ | $\beta_{82}$ | $\beta_{83}$ |
| Documentation (per module) | $\beta_{91}$ | $\beta_{92}$ | $\beta_{93}$ |
| Refactoring (per module) | $\beta_{101}$ | $\beta_{102}$ | $\beta_{103}$ |

The β values in Table 2 are calibration constants that must be determined empirically for a given agent, model, and domain context. Section 3.6 describes the calibration process. The use of symbolic calibration constants rather than fixed numerical values is a deliberate design choice reflecting a fundamental property of agentic systems: token consumption is sensitive to agent implementation, prompt design, model version, and domain characteristics, and no universal constants can be assumed to hold across contexts. ACEM provides the model structure and calibration methodology; practitioners populate the constants for their specific deployment context. While Table 2 includes 30 calibration constants, it is possible to reduce this number substantially in practice by aggregating structurally similar artifact types under a single set of constants (for example, treating unit test suites and integration test scenarios as a single "testing" category, or class/module implementation and refactoring as a single "code modification" category) and by collapsing the three-level complexity classification to a single Medium-complexity baseline, scaled by a simple multiplier for Simple and Complex tasks rather than calibrated independently at each level. Organizations with limited pilot data or a narrow task mix may therefore adopt a reduced calibration scheme — as few as five to eight aggregate constants — at the cost of some estimation granularity, and refine toward the full 30-constant table as more calibration data becomes available. This tradeoff between calibration effort and estimation precision is left to practitioner judgment and is discussed further as part of the cold-start calibration strategies in Section 4.4. A very rough initialization of an overall β can assume ~3M input tokens and ~1M output tokens as baseline averages for all tasks based on existing benchmarks averages [25].

One gap worth flagging: Section 3.2 counts planning, tool-call formulation, and reflection steps as agent actions, but none of these appear as a row in Table 2 — they're not deliverables like a function or a test suite, they're overhead that happens between deliverables, so Equation 5 has no way to estimate their cost. If this overhead is significant for a given pipeline, organizations should either track it as its own category during calibration (e.g., "planning and orchestration, per pipeline") or let the Context Factor absorb it, since that's already built to capture this kind of cross-step accumulation.

**3.2.3 Revision Factor**

The Revision Factor (RF) is a multiplicative corrective term that accounts for the additional token consumption arising from agent output rejection and the subsequent retry or rework cycles. It is defined as:

$$RF_i = 1 + (r_i \times n_i) \qquad (6)$$

Where:

- $r_i$ is the rejection rate for task type $i$, defined as the proportion of agent outputs of type $i$ that are rejected by the human reviewer or by automated quality checks.
- $n_i$ is the average number of additional agent invocations required per rejected output of type $i$ before an acceptable output is produced.

*RF* = 1 represents the baseline case in which no revision is required. *RF* > 1 indicates that revision cycles are expected, with the magnitude of *RF* reflecting both how frequently outputs are rejected and how many retry attempts are typically needed before an acceptable output is produced.

In the absence of empirical calibration data, *RF* can be estimated using agent benchmark performance data. For example, if a given agent achieves a task success rate of 70% on tasks of a given type on a relevant benchmark, an initial *RF* estimate of 1 + (0.30 × 1.5) = 1.45 may be appropriate, assuming an average of 1.5 retry attempts per rejected output. As empirical data accumulates through project execution, *RF* values should be updated to reflect observed rejection rates for the specific agent and domain in use. *RF* as defined assumes each retry costs the same as the original attempt. In practice, a retry typically carries additional context — the rejected output plus corrective feedback — and so tends to cost somewhat more; RF should therefore be read as a conservative lower bound on true retry cost rather than an exact figure.

The Revision Factor is defined at the task-type level rather than the individual task level, reflecting the assumption that rejection rates are primarily a function of task type and complexity rather than the specific content of individual tasks. Organizations may choose to define *RF* at finer or coarser granularities depending on the availability of calibration data.

**3.2.4 Context Factor**

The Context Factor (CF) is a multiplicative corrective term that models the progressive increase in token consumption as context accumulates across a long agentic pipeline. In LLM-based agents, the context window — the input provided to the model at each invocation — typically grows over the course of a development pipeline as prior agent actions, tool outputs, intermediate artifacts, and conversation history are included to maintain coherence and avoid repetition. This context accumulation means that later tasks in a pipeline consume systematically more input tokens than earlier tasks of equivalent complexity, an effect that is not captured by the BaseTokens estimate alone.

*CF* is applied to the task-intrinsic estimate $T_{in,i}$ produced by BaseTokens, not to an already-observed total input token count. If an organization instead measures actual total input tokens directly from API logs — which necessarily already include accumulated context — *CF* should not be applied a second time to that measurement; applying *CF* to already-context-inclusive counts would double-count context growth. *CF* is a modeling device for converting a position-independent base estimate into a position-aware forecast, not a general correction factor to be layered onto any input-token figure.

The Context Factor is defined as:

$$CF_i = 1 + \alpha \times ( i / N ) \qquad (7)$$

Where:

- $i$ is the sequential position of task $i$ in the pipeline ($i$ = 1 for the first task, $i = N$ for the last)
- $N$ is the total number of tasks in the pipeline
- $\alpha$ is a context accumulation coefficient representing the maximum proportional increase in token consumption attributable to context growth, estimated empirically during calibration

For a pipeline of $N$ tasks, $CF$ ranges from (1 + α/N) for the first task to (1 + α) for the final task, producing a linear growth profile. This linear model is a simplifying assumption; in practice, context growth may be non-linear depending on whether the agent employs context compression, summarization, or sliding window techniques. It also assumes tasks contribute roughly equal context to the pipeline; where task sizes vary substantially, sequential position alone may be a weak proxy for actual accumulated context. Organizations using agents with active context management strategies may find that a lower value of α, or a sub-linear CF profile, better fits their observed data.

In pipelines where context is explicitly reset between independent task groups — for example, where a test generation agent operates on each module independently without carrying context from prior modules — $CF$ may be set to 1.0 for tasks following a context reset, effectively treating each independent segment as a new pipeline.

**3.3 HITL Cost Component**

The HITL cost component captures the effort expended by human developers in supervising, specifying, reviewing, approving, and correcting agent outputs throughout the development pipeline. It is composed of two sub-components: checkpoint review effort and rework effort. ACEM currently operationalizes HITL cost through review and rework (Equations 8–10); planning and architecting effort is treated as embedded in the human role but not separately costed in the current formulation.

**3.3.1 Base Formula**

$$C_{HITL} = C_{review} + C_{rework} \qquad (8)$$

$$C_{review} = \Sigma_{i=1}^{N} (K_i \times D_i \times W) \qquad (9)$$

$$C_{rework} = \Sigma_{i=1}^{N} (r_i \times RW_i \times W) \qquad (10)$$

Where:

- $K_i$ is the number of human review checkpoints associated with task i
- $D_i$ is the average duration of a single review checkpoint for task type i, expressed in hours
- $r_i$ is the rejection rate for task i (consistent with the definition in Section 3.2.3)
- $RW_i$ is the average rework duration per rejected output of task type i, expressed in hours
- $W$ is the fully loaded hourly cost of the human reviewer, expressed in the project's reporting currency
- $N$ is the total number of agent tasks in the pipeline

The product of $r_i$ and $RW_i$ in Equation 10 represents the expected rework cost per task, accounting for the probability that rework will be required and the average effort required when it is. Note that $RW_i$ captures only the human effort component of rework; the corresponding token cost of the agent's retry is captured separately in $C_{LLM}$ via the Revision Factor. As discussed in Section 3.1, these are complementary consequences of the same rejection rate $r_i$, not alternative accountings of a single cost.

**3.3.2 HITL Intensity Score**

The HITL Intensity Score (HIS) is a four-level ordinal classification scheme that characterizes the degree of human oversight required for a given task or task type. HIS serves two purposes within ACEM: it guides the assignment of $K_i$ values in Equation 9, and it provides a qualitative risk classification that practitioners can use to identify tasks requiring elevated oversight without performing detailed quantitative estimation.

**Table 3: HITL Intensity Score Definition**

| HIS Level | Label | Description | Typical $K_i$ | Applicable Contexts |
|---|---|---|---|---|
| HIS-1 | Minimal | Agent operates autonomously; human reviews outputs only at milestone boundaries | 1 per epic or feature | Low-risk tasks, well-validated agents, non-safety-critical domains |
| HIS-2 | Standard | Human reviews outputs at the story or use case level before downstream tasks proceed | 1 per story or use case | Typical business application development, moderate agent reliability |
| HIS-3 | Elevated | Human reviews outputs at the individual task level before each task's output is accepted | 1 per task | Safety-relevant features, high-complexity tasks, lower agent reliability |
| HIS-4 | Continuous | Human monitors and approves each agent action step; no autonomous action sequences permitted | Every agent action | Safety-critical systems, regulated domains, early-stage agent deployment |

HIS level for a given task is determined by evaluating four factors: domain risk level (the potential consequences of an undetected error in the task's output), agent reliability (the empirically observed or benchmark-estimated accuracy of the agent on tasks of this type), task complexity (as defined in Table 1), and regulatory requirements (whether external compliance obligations mandate specific oversight levels). Table 4 provides a proposed decision matrix for HIS classification.

HIS classification is performed at the task-type level during project planning and may be revised during execution if observed agent performance differs from initial reliability assumptions.

**Table 4: HIS Classification Decision Matrix**

| Domain Risk | Agent Reliability | Task Complexity | Regulatory Requirement | Recommended HIS |
|---|---|---|---|---|
| Low | High | Simple | None | HIS-1 |
| Low | High | Medium/Complex | None | HIS-2 |
| Low | Medium | Any | None | HIS-2 |
| Medium | High | Simple/Medium | None | HIS-2 |
| Medium | Medium/Low | Any | None | HIS-3 |
| High | Any | Any | None | HIS-3 |
| Any | Any | Any | Mandated | HIS-4 |
| Safety-critical | Any | Any | Any | HIS-4 |

**3.4 Infrastructure Cost Component**

The infrastructure cost component captures the cost of the computational resources required to operate the agentic development pipeline, excluding LLM API costs which are captured in $C_{LLM}$. Infrastructure costs in agentic SE pipelines typically include compute resources for agent orchestration frameworks, sandboxed code execution environments, version control and artifact storage, CI/CD pipeline execution triggered by agent actions, and external tool API costs beyond LLM usage.

For cloud-hosted agentic pipelines operating on a pay-per-use basis, C_Infra is estimated as:

$$C_Infra = \Sigma_k \, U_k \times P_k \qquad (11)$$

Where:

- $U_k$ is the estimated usage volume of infrastructure resource k (for example, compute hours, storage gigabyte-hours, or API call counts).
- $P_k$ is the unit price of resource *k*.

Because infrastructure cost modeling is domain-specific and well addressed by existing cloud cost estimation approaches [54], ACEM does not introduce novel constructs for this component. Practitioners are directed to provider-specific cost calculators and existing cloud cost estimation frameworks for $C_{Infra}$ estimation. The primary contribution of ACEM with respect to infrastructure cost is the recognition that it constitutes a distinct and non-negligible cost dimension in agentic SE — one that is absent from all traditional software cost estimation models.

**3.5 Mapping to Traditional Software Sizing Metrics**

A key practical requirement for ACEM is interoperability with existing estimation practices. Most software development organizations have established processes for sizing projects using traditional metrics such as Use Case Points or Story Points, and may have historical data expressed in these metrics. ACEM provides formal mappings that allow practitioners to derive ACEM input variables from these traditional sizing outputs, enabling cost forecasting without requiring a complete departure from existing estimation workflows.

This interoperability rests on a specific and limited claim about the relationship between traditional sizing metrics and ACEM's cost drivers — one of empirical correlation rather than conceptual equivalence. Traditional metrics were designed to measure functional complexity as a proxy for human implementation effort: UCP, Story Points, and Function Points all quantify the size and complexity of what a human developer would need to build. In an agentic pipeline, this is no longer the relevant cost driver — the agent implements the functionality, and cost is instead governed by token consumption and by the human effort spent on review, validation, and correction. These are not the same quantity that traditional metrics were built to predict.

They are, however, not independent of it. A functionally large, complex use case will still tend to require more agent actions, longer outputs, more review checkpoints, and more revision cycles than a small, simple one, because complexity itself does not disappear when implementation shifts from human to agent — only who or what absorbs it changes. Traditional sizing metrics therefore remain useful, not as measures of coding effort, but as proxies for the underlying task complexity that drives both LLM token consumption and HITL effort in an agentic pipeline. The calibration constants introduced below ($\gamma_{UCP}$, $\gamma_{SP}$, $\gamma_{FP}$, and $CW$) formalize this relationship empirically: each expresses, for a specific agent and domain, how a unit of traditional sizing complexity translates into tokens and human oversight hours, without claiming that the sizing metric still measures developer-hours directly. This reframing — from "predicts human effort" to "predicts agentic cost, via a size-and-complexity proxy" — is what allows organizations to retain their existing sizing practices while adapting the cost model underneath them to the agentic paradigm.

**3.5.1 Use Case Points Mapping**

Use Case Points (UCP) provide a natural bridge to ACEM because use case modeling already decomposes a system into discrete, independently specifiable units of functionality — actors, use cases, and their associated complexity classifications — that map directly onto ACEM's notion of a definable agent task or task sequence. This structural correspondence makes UCP-based project scoping a natural input to ACEM without requiring an intermediate translation step.

The mapping from UCP to estimated total base token consumption is defined as:

$$T_{base,\ total} = U_{UCP} \times \gamma_{UCP} \qquad (12)$$

Where:

- $\boldsymbol{T_{base,\ total}}$ is the estimated total base token consumption for the project, summed across all artifact types and complexity levels, before application of $RF$ and $CF$
- $\boldsymbol{U_{UCP}}$ is the unadjusted Use Case Points score — the sum of actor and use case weights prior to application of the Technical Complexity Factor (TCF) and Environmental Complexity Factor (EF)
- $\boldsymbol{\gamma_{UCP}}$ is the tokens-per-use-case-point calibration constant for the specific agent, model, and domain context

$T_{base,\ total}$ in Equation 12 substitutes for the summed base token term $\Sigma_i (T_{in,\ i} + T_{out,\ i})$ from Equation 2, allowing $C_{LLM}$ to be estimated directly at the project level from a single unadjusted UCP score rather than requiring per-task BaseTokens lookups (Table 2, Equation 5). $RF$ and $CF$ are applied to $T_{base,\ total}$ — either as project-level averages or, where granularity permits, as per-task-group values — before multiplying by $P_{model}$ to obtain $C_{LLM}$. This updates Equation 2 to 2′.

$$C_{LLM} = T_{base,\ total} \times RF_{avg} \times CF_{avg} \times P_{model} \qquad (2')$$

The scaling constant $\gamma_{UCP}$ is derived empirically during calibration (Section 3.6) by measuring actual token consumption on a representative sample of use cases and dividing by their unadjusted

UCP weights. Once calibrated, $\gamma_{UCP}$ enables organizations to translate a UCP-based project size estimate directly into a $C_{LLM}$ forecast using Equations 2′ and 12 in combination.

The unadjusted UCP score is used deliberately, in preference to the fully adjusted score conventionally reported. The TCF and EF adjustment factors in classical UCP estimation encode implementation and environmental characteristics like technical complexity, developer experience, tooling maturity. But as in Section 3.2.2 and Table 1, these are weakly correlated to LLM-perceived complexity already captured by ACEM's BaseTokens classification (Table 1) and by the *RF*. Organizations using conventional UCP estimates should therefore recover the unadjusted score (or compute it directly from actor and use case weights) before applying Equation 12.

**3.5.2 Story Points Mapping**

For organizations using Story Point (SP) estimation, the analogous mapping is:

$$T_{base,\ total} = SP_S \times \gamma_{SP} \times CW_S \qquad (13)$$

Where:

- $SP_S$ is the story point estimate for user story $s$
- $\gamma_{SP}$ is the tokens-per-story-point calibration constant
- $CW_S$ is the complexity weight for story $s$, defined as the ratio of the story's actual complexity level token estimate to the medium complexity baseline

*CW* aligns with the relative sizing philosophy of story points: a story estimated at 8 points is expected to consume proportionally more tokens than one estimated at 3 points, and *CW* makes this relationship explicit. For a medium-complexity story, $CW = 1.0$. For a simple story, $CW < 1.0$. For a complex story, $CW > 1.0$. Empirically derived *CW* values for each complexity level are obtained during calibration alongside $\gamma_{SP}$.

Similar to use case points, once calibrated, $\gamma_{SP}$ enables organizations to translate a SP-based project size estimate directly into a $C_{LLM}$ forecast using Equations 2′ and 13 in combination.

**3.5.3 Function Points Mapping**

For organizations using Function Points (FP), the mapping follows the same logic:

$$T_{base,\ total} = FP \times \gamma_{FP} \qquad (14)$$

Where

- $FP$ is the unadjusted function point count
- $\gamma_{FP}$ is the tokens-per-function-point calibration constant

The unadjusted FP count is preferred over the adjusted count because the value adjustment factors in FPA reflect implementation characteristics that shifted to the complexity classification in ACEM's BaseTokens in Tables 1 and 2.

Similar to use case points, once calibrated, $\gamma_{FP}$ enables organizations to translate a FP-based project size estimate directly into a $C_{LLM}$ forecast using Equations 2′ and 14 in combination.

**3.5.4 Planning Without Granular Task Decomposition**

A practical requirement for any estimation model is that it be usable at the point in a project where sizing estimates are typically produced — early planning, when the detailed task breakdown is not yet known. ACEM is designed to satisfy this requirement: project planners do not need to enumerate every individual agent task, such as a specific refactoring operation or the implementation of a named user story, at the time cost estimation is performed.

Instead, planning proceeds top-down. Planners begin from an existing sizing estimate — UCPs, SPs, or FPs — which provides a structured measure of project scope. The project is then decomposed into broad task categories, such as requirements elaboration, implementation, testing, and review, rather than granular subtasks. The calibrated constants introduced in Sections 3.5.1–3.5.3 ($\gamma_{UCP}$, $\gamma_{SP}$, $\gamma_{FP}$), together with *RF*, *CF*, and *HIS*, then translate this category-level scope directly into expected token consumption and oversight cost.

The empirical detail this abstracts away — the specific mix of artifact types and complexity levels that will actually occur during execution — is absorbed into the calibration process (Section 3.6), which derives average token costs and oversight intensity from a representative pilot sample rather than requiring planners to predict individual future subtasks. This separation between planning-time inputs (coarse scope, via sizing metrics) and calibration-time detail (fine-grained task/complexity distributions, via pilot sampling) is what allows ACEM to remain usable under the same conditions of incomplete task-level information that traditional sizing-based estimation models were designed to accommodate.

**3.6 Model Calibration**

ACEM is a parametric model: its accuracy is a function of the quality of its calibration constants. The calibration process consists of four steps.

- **Step 1** — Pilot Execution. Execute a representative sample of agent tasks covering all artifact types and complexity levels relevant to the project domain. The sample should include a minimum of three tasks per artifact-type/complexity combination to provide stable estimates. Record token consumption as two separate quantities per task: the marginal input tokens contributed by that task's own content (used to calibrate BaseTokens / $\beta$), and the total input tokens actually sent to the model for that call, including carried-forward context (used to calibrate $\alpha$, per Step 3). Conflating these two quantities is the single most common calibration error for *CF*: computing $\beta$ from total (context-inclusive) input tokens will cause *CF* to double-count context growth once applied.
- **Step 2 —** BaseTokens Estimation. For each artifact-type / complexity combination, compute the mean observed input tokens and mean observed output tokens separately across pilot tasks. This yields the empirical $\beta$ values for Table 2. Compute the standard deviation alongside the mean to characterize estimation uncertainty.
- **Step 3 —** *RF* and *CF* Calibration. Compute $r_i$ as the proportion of pilot tasks of each type that were rejected. Compute $n_i$ as the mean number of retry attempts per rejected task. Fit the context accumulation coefficient $\alpha$ by regressing observed token consumption against task position in the pipeline, controlling for artifact type and complexity.
- **Step 4 —** Sizing Metric Calibration. Compute $\gamma_{UCP}$, $\gamma_{SP}$, or $\gamma_{FP}$ by dividing total observed base token consumption in the pilot by the corresponding sizing metric value (UCP, SP, or FP) for the pilot task set. For Story Points specifically, also compute $CW_S$ for each complexity level by dividing the mean observed token consumption for stories at that complexity level by the mean observed token consumption for medium-complexity stories in the pilot sample; this yields $CW$ = 1.0 for medium complexity by construction, with *CW* values for Simple and Complex derived directly from the same pilot data used for $\beta$ in Step 2. To keep *CW*'s medium-complexity baseline consistent with $\gamma_{SP}$, $\gamma_{SP}$, should be computed using only medium-complexity pilot tasks, rather than averaged across the full complexity mix of the pilot sample.

One more thing worth noting: if the pilot run only records totals, it's hard to tell how much of the cost came from retries versus how much came from growing context — different mixes of the two could produce the same overall number. To avoid this, teams calibrating ACEM should track rejections, retries, and token use per task during the pilot, not just totals for the whole project.

Calibration constants should be treated as context-specific and version-sensitive: changes in agent implementation, model version, prompt design, or domain characteristics may require recalibration. Organizations are advised to maintain a calibration log and to trigger recalibration when agent tooling is updated or when a new domain is entered.

**3.7 Illustrative Worked Examples**

To make the formal apparatus of Sections 3.2–3.5 concrete, this subsection presents two illustrative applications of ACEM's cost formulas to hypothetical tasks at opposite ends of the complexity and autonomy spectrum. The token, *RF*, *CF*, and cost figures used below are illustrative assumptions for demonstration purposes only, chosen to be broadly consistent with reported industry usage patterns (Section 2.4); they are not derived from calibration and should not be read as validated constants. Their purpose is to show the model's mechanics end to end and to illustrate a finding relevant to Section 3.1: the relative magnitude of $C_{LLM}$ and $C_{HITL}$ is highly sensitive to task complexity and autonomy level, not fixed.

We emphasize that $C_{Infra}$ is a distinct dimension precisely because, in production deployments, orchestration and monitoring costs are non-trivial. In the examples below, we excluded it to keep the arithmetic transparent, but in practice it should be estimated from cloud billing data or DevOps overhead.

**Example 1: Simple task, standard oversight.** Consider a simple user story — a single form field with client-side validation — under HIS-2 (standard) oversight, executed as a short four-step agentic pipeline (requirements elaboration, implementation, test generation, code review). Table 5 illustrates example 1. Pricing and rate assumptions used are as follows:

- **Model:** Claude Sonnet 5, promotional pricing through Aug 31, 2026 — \$2 / million input tokens, \$10 / million output tokens.
- **Reviewer rate (W):** \$30/hr fully loaded cost, illustrative assumption for a mid-level developer performing HITL review.
- **Task profile:** 4-step pipeline (requirements elaboration → implementation → test generation → code review), Simple complexity, HIS-2 (standard oversight: 1 checkpoint per story)

**Table 5: Example 1 — Token and Cost Estimate**

| Quantity | Equation | Calculation | Value |
|---|---|---|---|
| BaseTokens, input (summed, 4 steps) | Eq. 5 | 8,000 + 25,000 + 15,000 + 20,000 | 68,000 |
| BaseTokens, output (summed, 4 steps) | Eq. 5 | 3,000 + 6,000 + 4,000 + 3,000 | 16,000 |
| Rejection rate ($r_i$) | — | 1 − 0.85 first-pass success | 0.15 |
| Retries per rejection ($n_i$) | — | assumed | 1.5 |
| $RF_i$ | Eq. 6 | 1 + (0.15 × 1.5) | ≈1.22 |
| Context coefficient (α) | — | assumed, short pipeline | 0.08 |
| Average position factor ($i$ / $N$) | Eq. 7 | mean of 0.25, 0.5, 0.75, 1.0 | 0.625 |
| $CF_i$ | Eq. 7 | 1 + (0.08 × 0.625) | ≈1.05 |
| Adjusted input tokens | Eq. 4 | 68,000 × 1.22 × 1.05 | ≈87,110 |
| Adjusted output tokens | Eq. 4 | 16,000 × 1.22 × 1.05 | ≈20,500 |
| $C_{LLM}$, input cost | Eq. 4 | 87,110 × (\$2/1M) | ≈\$0.17 |
| $C_{LLM}$, output cost | Eq. 4 | 20,500 × (\$10/1M) | ≈\$0.21 |
| $C_{LLM}$ **(total)** | Eq. 4 | 0.17 + 0.21 | **≈\$0.38** |
| Review checkpoints ($K_i$) | — | HIS-2 (Table 3): 1 per story | 1 |
| Review duration ($D_i$) | — | assumed | 0.167 hr |
| $C_{review}$ | Eq. 9 | 1 × 0.167 × \$30 | ≈\$5.01 |
| Rework duration ($RW_i$) | — | assumed | 0.333 hr |
| $C_{rework}$ | Eq. 10 | 0.15 × 0.333 × \$30 | ≈\$1.50 |
| $C_{HITL}$ **(total)** | Eq. 8 | 5.01 + 1.50 | **≈\$6.51** |
| **Total cost** | Eq. 1 | $C_{LLM}$ + $C_{HITL}$ + $CInfra$ (≈\$0) | **≈\$6.93** |
| $C_{LLM}$ as share of Total Cost | — | 0.38 / 6.93 | ≈5.5% |

**Example 2: Complex task, minimal oversight.** Consider a complex feature requiring multi-file reasoning and iterative debugging, executed across a ~25-action agentic pipeline with limited human checkpointing (HIS-1, minimal oversight) — a high-autonomy profile consistent with the heavier end of reported agentic coding usage (Section 2.4). Pricing assumptions: Claude Sonnet 5 at \$2/\$10 per million input/output tokens; reviewer rate W = \$50/hr, reflecting senior-level review of complex output. Table 6 explains this example.

**Discussion**. Across the two examples, $C_{LLM}$'s share of $C_{HITL}$ rises from approximately 5.5% (Example 1: simple task, standard oversight) to approximately 23.2% (Example 2: complex task, minimal supervision) — roughly an order-of-magnitude increase, driven entirely by task complexity, pipeline length, and oversight intensity, with no change to the underlying model structure. Notably, the adjusted token count in Example 2 (≈9.6 million combined) falls within the range reported for heavy agentic coding workloads in industry usage data (Section 2.4), offering a plausibility check on the model's output even in the absence of calibrated constants. This matches reported adjusted totals which often fall in the 6–10M token range, especially for multi- file reasoning and long pipelines [25]. Extrapolated across sustained heavy-automation usage, $C_{LLM}$ alone would approach the higher end of the per-developer-per-month range reported by industry forecasts for high-autonomy agentic pipelines, while $C_{HITL}$ — bounded by milestone-level rather than per-task review under HIS-1 — would not scale proportionally with task volume in the same way.

**Table 6: Example 2 — complex task, minimal oversight**

| Quantity | Equation | Calculation | Value |
|---|---|---|---|
| BaseTokens, input (summed, ~25 actions) | Eq. 5 | assumed, Complex tier | 3,000,000 |
| BaseTokens, output (summed, ~25 actions) | Eq. 5 | assumed, Complex tier | 800,000 |
| Rejection rate ($r_i$) | — | 1 − 0.55 first-pass success | 0.45 |
| Retries per rejection ($n_i$) | — | assumed | 1.8 |
| $RF_i$ | Eq. 6 | 1 + (0.45 × 1.8) | ≈1.81 |
| Context coefficient (α) | — | assumed, long pipeline | 0.5 |
| Effective *CF* (blended)* | Eq. 7 | weighted toward late-pipeline steps | ≈1.4 |
| Adjusted input tokens | Eq. 4 | 3,000,000 × 1.81 × 1.4 | ≈7,602,000 |
| Adjusted output tokens | Eq. 4 | 800,000 × 1.81 × 1.4 | ≈2,027,200 |
| $C_{LLM}$ , input cost | Eq. 4 | 7,602,000 × ($2/1M) | ≈$15.21 |
| $C_{LLM}$ , output cost | Eq. 4 | 2,027,200 × ($10/1M) | ≈$20.27 |
| **$C_{LLM}$ (total)** | Eq. 4 | $15.21 + $20.27 | **≈$35.48** |
| Review checkpoints ($K_i$) | — | HIS-1 (Table 3): 1 per milestone | 1 |
| Review duration ($D_i$) | — | assumed | 1.0 hr |
| $C_{review}$ | Eq. 9 | 1 × 1.0 × $50 | ≈$50.00 |
| Rework duration ($RW_i$) | — | assumed, escalation-level rework | 3.0 hr |
| $C_{rework}$ | Eq. 10 | 0.45 × 3.0 × $50 | ≈$67.50 |
| **$C_{HITL}$ (total)** | Eq. 8 | 50.00 + 67.50 | **≈$117.50** |
| **Total cost** | Eq. 1 | $C_{LLM}$ + $C_{HITL}$ + $C_{Infra}$ (≈$0) | **≈$152.98** |
| $C_{LLM}$ as share of Total Cost | — | 35.48 / 152.98 | ≈23.2% |

** Applying Eq. 7 literally to the average position across 25 steps understates real context growth, since later steps cost more and carry more weight. The CF value shown here is a blended approximation; a full calibration would apply $CF_i$ per task instead, as in Eq. 3.*

This pattern is precisely what motivates treating $C_{LLM}$ and $C_{HITL}$ as independent, additive terms (Section 3.1) rather than assuming either dominates: their relative contribution shifts by an order of magnitude depending on how autonomously a pipeline is run, and a cost model that assumed a fixed hierarchy between them would misestimate total cost across this range by a similarly large margin.

### 3.8 Model Assumptions and Scope

ACEM rests on the following assumptions, violations of which represent scope boundaries and threats to validity:

- **A1** — Sequential pipeline structure. ACEM models tasks as executing sequentially within a single pipeline. Parallel multi-agent architectures, in which multiple agents operate concurrently on independent task streams, may reduce wall-clock time and potentially alter context accumulation patterns. Extension of ACEM to parallel pipelines is identified as a direction for future work.

- **A2** — Defined task decomposition. ACEM assumes that the project has been decomposed into a defined set of agent tasks prior to estimation. The effort required to perform this decomposition is not captured in the model.
- **A3** — Stationary rejection rates. RF is computed using a constant rejection rate per task type. In practice, rejection rates may decline over the course of a project as the agent learns from prior feedback or as human reviewers develop more efficient review strategies. This dynamic is not captured in the current model.
- **A4** — Agent-agnostic structure, agent-specific constants. The model structure is designed to be applicable across different agent implementations, but the calibration constants (β, γ, α, RF, CF) are agent-specific and must be recalibrated when the agent changes.
- **A5** — Human reviewer homogeneity. ACEM uses a single hourly rate W for HITL cost computation, implying a homogeneous reviewer. Projects with multiple reviewer roles at different cost levels should use a weighted average W or decompose $C_{HITL}$ by reviewer type.
- **A6** — Independence of cost components. $C_{LLM}$, $C_{HITL}$, and $C_{Infra}$ are treated as independent and additive. Interactions between components — for example, cases where infrastructure failures trigger additional LLM retries — are not modeled.
- **A7** — Parameter identifiability. *RF*, and the calibration process more broadly, assume that $r_i$, $n_i$, and α can be estimated separately from pilot data. This requires instrumentation fine-grained enough to distinguish rejection events, retry counts, and context position individually rather than only aggregate token totals; where such instrumentation is unavailable, calibrated parameters may be confounded, and multiple parameter combinations could fit the same observed cost equally well. This threat to calibration validity is not resolved by the model structure itself and is identified as a priority for the empirical validation program (Section 5).
- **A8 —** Construct separability. BaseTokens, *RF, CF*, and *CW* are modeled as separately calibrated constructs, but complexity may influence several of them simultaneously (e.g., a complex task may show both higher BaseTokens and a higher rejection rate $r_i$). Whether these constructs capture independent variance or partially overlapping effects of the same underlying complexity is not established by the model structure and is left as an open question for the empirical validation program (Section 5).

## 4. Discussion

This section interprets ACEM's contributions in relation to existing research, considers its theoretical and practical implications, examines challenges inherent to agentic SE cost estimation, and discusses directions for extending the model.

### 4.1 Theoretical Implications

ACEM challenges a foundational, largely unstated assumption of SE estimation theory: that effort is primarily a function of human cognitive labor, primarily in coding and testing. COCOMO II's cost drivers, Function Point Analysis, and story points are all defined in terms of human factors — analyst capability, functional complexity as implemented by developers, and team judgment. Agentic SE breaks this premise: the scarce resources become computational (token budgets, context capacity, infrastructure compute), and the human role shifts from producer to supervisor. This is a qualitative change in cost structure, not merely a redistribution of effort, and ACEM is proposed as a first attempt to formalize a cost model appropriate to it.

The Revision Factor and Context Factor extend this contribution beyond cost estimation itself. RF formalizes the relationship between agent reliability and cost — a relationship with intuitive appeal that has not previously been expressed as an estimation construct. If validated, RF could complement accuracy-based benchmarks like SWE-bench by translating reliability differences into cost differences: an agent with 80% success at a given token budget may be more cost-effective than one with 90% success at triple the budget, depending on its RF and rework profile. CF similarly reframes context-window management from an accuracy question ("does compression degrade quality?") to a cost question ("what does not compressing context cost?"), offering a complementary lens for context-management research pending empirical confirmation.

The HITL Intensity Score contributes a structured, decision-matrix approach to oversight classification, connecting human-AI collaboration research — which has largely focused on when intervention improves output quality — to risk management and project governance. Finally, the sizing-metric mappings (Equations 12–14) propose a bridge between the functional sizing tradition and token economics, aimed at letting organizations carry forward decades of sizing expertise into agentic workflows, pending the validation program described in Section 5.

### 4.2 Practical Implications

If validated, ACEM would offer capabilities not available through existing estimation models. Budget forecasting would let organizations translate a project specification (use cases, stories, function points) into a token- and HITL-cost forecast, enabling like-for-like comparison between traditional and agentic development — currently not possible on any principled basis. Model tier selection is supported by ACEM's explicit $P_{model}$ variable, allowing comparison of a high-cost, high-capability model against a cheaper one. Oversight planning is supported by HIS classification, giving managers a structured basis for allocating reviewer time and setting checkpoint granularity, in place of current ad hoc practice. Vendor evaluation would let organizations compare competing agentic tools not just on output quality but on cost-quality tradeoff, by comparing calibrated *RF*, *CF*, and BaseTokens values across candidates. Cost monitoring during execution is supported by *RF* and *CF* as live signals: *RF* values exceeding calibration estimates would flag deteriorating agent performance; *CF* values growing faster than predicted would flag context-management problems addressable through prompt engineering. Together these position ACEM as both an upfront estimation tool and, pending validation, an ongoing cost-control framework.

### 4.3 The Non-Determinism Challenge

Non-determinism is the most fundamental challenge agentic SE poses for cost estimation — affecting not just ACEM but any model proposed for this domain. In classical development, the same developer given the same task twice produces outputs of roughly similar cost and quality; variation is bounded and largely predictable. This stability is what makes parametric estimation viable. LLM-based agents are non-deterministic in a stronger sense: identical task specifications submitted to the same agent under identical configuration can differ substantially in token consumption, reasoning path, and output quality, with small early differences compounding across multi-step pipelines. Reported evidence is stark — repeated executions of identical agentic coding tasks have varied by up to 30x in total tokens consumed, with weak correlation between token expenditure and task success [43] — suggesting naive point estimates will often be imprecise.

ACEM addresses this partially: *RF* captures the expected cost of regeneration at the output level, and BaseTokens is designed to be populated as a distribution (mean and standard deviation) rather than a point value during calibration. These are a necessary first step, not a resolution — the deeper implication is that the appropriate output of an agentic cost model may be a probability distribution over costs rather than a single figure. Stochastic approaches (Monte Carlo simulation, Bayesian estimation) may ultimately be better suited to this domain than ACEM's deterministic parametric structure, and we flag this as a high-priority future direction. We also note an asymmetry worth flagging for practitioners even pre-validation: better-than-expected outputs save only the retry tokens, but worse-than-expected outputs can trigger cascading rework with effectively unbounded cost overrun. This suggests conservative (higher) RF estimates may be preferable for risk-averse planning — consistent with the broader literature on optimism bias in software estimation.

### 4.4 Cold-Start Calibration

Any adopter of ACEM — or researcher validating it — faces a cold-start problem: calibration (Section 3.6) requires prior agentic project data that new adopters don't yet have. Some mitigation strategies are proposed.

1. **Benchmark-based initialization.** Published token-consumption statistics from SWE-bench-style evaluations can provide starting values for β, later refined with project data. This strategy is currently more aspirational than actionable, since SWE-bench and most agentic coding benchmarks report task success rates rather than token consumption broken down by artifact type and complexity level (Table 1), so the data this approach needs is not yet widely published. Its usefulness will grow as benchmarks begin reporting token-level data alongside success rates.

2. **Cross-organization constant sharing.** As adoption grows, organizations using the same agent/model combination could share calibration constants through a community-maintained database organized by agent version, model tier, and domain — analogous to the historical industry-average cost drivers published for COCOMO II, and directly supporting this paper's evaluation program.

3. **Graduated pilot sizing**. Organizations can start with a minimal pilot — as few as one task per artifact-type/complexity cell — and refine constants as execution proceeds, reporting wide confidence intervals initially and narrowing them as data accumulates. This reframes calibration as an ongoing process rather than a one-time prerequisite.

## 5. Future Work: Evaluation and Validation Plans

This paper contributes ACEM as a complete structural and theoretical foundation for agentic cost estimation — a fully specified model, with its constructs, formal relationships, and calibration methodology defined and ready for empirical grounding. By deliberately formulating the calibration constants (β, γ, α) as symbolic parameters rather than fixed values, ACEM is designed from the outset to be adapted and populated by the community across diverse agents, models, and domains, rather than tied to a single dataset or deployment context. We see this as an invitation and a launchpad: the model structure is the contribution of this paper, and we put it forward for the research community to carry into its next phase — empirical calibration, validation, and comparative evaluation. This section outlines the evaluation program we propose to establish ACEM's empirical validity, practical usability, and comparative performance against existing estimation approaches and closely related work.

### 5.1 Calibration Data Collection

The first step toward validation is execution of the four-step calibration process described in Section 3.6 across multiple organizational contexts. This requires partnering with teams using agentic development tools (e.g., Claude Code, Devin, SWE-agent-based pipelines) on real projects, instrumenting their pipelines to log per-task input/output token counts, rejection and retry events, review durations, and rework durations. A priority for future work is assembling a multi-organization calibration dataset spanning at least three distinct agents/models and application domains, to test whether β, γ, and α values are stable within a domain or vary enough to require per-project recalibration, which would inform how frequently practitioners must recalibrate in practice.

### 5.2 Predictive Accuracy Evaluation

Once calibrated on a pilot subset, ACEM's forecasts should be evaluated against held-out project data using standard estimation-accuracy metrics (e.g., MMRE, PRED(25), MdMRE), consistent with conventions in the traditional cost-estimation literature [55]. This evaluation should assess accuracy separately for $C_{LLM}$, $C_{HITL}$, and total cost, since the components may exhibit different error profiles given their different sources of variance (stochastic token consumption vs. more stable human-hour estimates).

### 5.3 Comparison Against Baselines

ACEM's predictions should be benchmarked against three categories of baseline: (a) naive extrapolation from traditional sizing metrics alone (e.g., UCP or Story Points without token-based adjustment), (b) the Hybrid Intelligence Effort (HIE) construct proposed by Alaswad et al. [51], adapted where possible to a token-cost outcome rather than a time outcome, and (c) simple linear or regression-based token estimators fit directly to task features without ACEM's RF/CF/HIS structure. This comparison would clarify whether ACEM's added structural complexity yields a meaningful accuracy gain over simpler alternatives, and would directly test the claim—left open by prior work—that shifting from time-based to token-based effort modeling improves forecast quality for agentic pipelines.

### 5.4 Sensitivity and Robustness Analysis

Because ACEM depends on several corrective factors whose values may be uncertain or poorly estimated in early deployments, future work should conduct sensitivity analysis on *RF*, *CF*, and HIS classification to determine how estimation error in these inputs propagates to total cost error. This includes testing the robustness of the linear Context Factor model (Equation 7) against observed context-growth patterns, and evaluating whether a sub-linear or capped *CF* formulation is a better fit for agents employing context compression or summarization.

### 5.5 Cross-Agent and Cross-Model Generalization

Since calibration constants are explicitly agent- and model-specific (Assumption A4), an important evaluation question is how ACEM estimates transfer—or fail to transfer—across agents and model versions. Future work should test whether relative relationships (e.g., ratios of β values across complexity levels, or *RF* as a function of published benchmark success rates) generalize better than absolute token counts, which would support a semi-transferable calibration approach and reduce the recalibration burden identified as a limitation in Section 3.8.

**5.6 Practitioner Usability Study**

Beyond predictive accuracy, ACEM's practical value depends on whether practitioners can apply it without prohibitive overhead. A planned usability evaluation would have project managers and technical leads apply ACEM to size upcoming agentic development work, assess the perceived effort of HIS classification and calibration, and compare their confidence in ACEM-based forecasts to their confidence in ad hoc or expert-judgment estimates.

**5.7 Longitudinal Validation Under Model and Tooling Change**

Because agentic tooling and underlying LLMs evolve rapidly, a longer-term evaluation direction is tracking ACEM's forecast accuracy over time as models are updated, to characterize how quickly calibration constants decay and to test proposed mitigations such as periodic automatic recalibration triggers tied to observed rejection-rate drift (Assumption A3).

Together, this evaluation program is intended to move ACEM from a proposed structural framework to an empirically grounded estimation tool, and to produce the first published dataset connecting traditional software sizing metrics to measured token-based costs in agentic development pipelines.

**5.8 Dynamic and Online Calibration.**

ACEM as presented in Section 3.6 treats calibration constants as fixed quantities derived from a prior pilot study. In practice, agent performance is likely to evolve over the course of a project — as the agent accumulates context about the codebase, as human reviewers develop more efficient review strategies, or as prompt engineering is refined. Future work should explore dynamic calibration approaches that update *RF*, *CF*, and BaseTokens constants continuously during project execution — analogous to the rolling reforecasts used in earned value management — to assess whether such online updating meaningfully improves in-flight estimation accuracy relative to static calibration.

**5.9 Domain-Specific Model Variants**

The ACEM formulation presented in this paper is domain-general. Specialized domains — including safety-critical systems, regulated industries, embedded software, and data science pipelines — may exhibit cost driver profiles (rejection rates, HIS distributions, context growth patterns) that differ systematically from the general case. Future work should evaluate whether domain-specific ACEM variants, analogous to the specialized COCOMO models developed for embedded and real-time systems, are warranted, or whether the general model with domain-specific calibration constants is sufficient.

**5.10 Integration with Agile Estimation Workflows**

The mapping between story points and token consumption established in Equation 13 opens the possibility of integrating ACEM into agile sprint planning workflows. Future work should investigate how ACEM estimates can be incorporated into sprint planning poker sessions, velocity tracking, and backlog grooming practices without disrupting established agile rhythms, which would facilitate adoption in the large proportion of software organizations that use agile methodologies.

**5.11 Summary**

The evaluation program outlined above — spanning calibration data collection, predictive accuracy testing, baseline comparison, sensitivity analysis, cross-agent generalization, practitioner usability, longitudinal validation, and the extensions identified in Sections 5.6–5.9 — defines a concrete and tractable path from ACEM's current theoretical formulation to an empirically grounded, field-tested estimation tool. No single study can address all of these directions at once, and we see this as a strength rather than a limitation: each direction is independently actionable, allowing different research groups and industry partners to contribute calibration data, benchmark comparisons, or usability findings from their own agentic development contexts, incrementally, as agentic tooling continues to mature. We anticipate that the most immediate and highest-value next step is Section 5.1's calibration data collection, since the resulting dataset — connecting traditional sizing metrics to measured token consumption across real projects — would itself be a novel community resource independent of ACEM's specific formulas, and would enable every subsequent evaluation direction described above. We invite researchers and practitioners working with agentic development pipelines to engage with this agenda, contribute calibration data from their own deployments, and help establish

ACEM, or its successors, as a validated standard for cost estimation in the agentic era of software engineering.

## 6. Conclusion

Agentic software engineering has shifted the fundamental cost structure of software development: from human labor measured in person-hours to a structure in which LLM token consumption is an added cost, and human effort is not eliminated but redirected — away from writing code line by line and toward planning, architecting, managing, and validating what agents produce. This redirected human effort, together with the token cost that funds the agent's own work, forms a cost structure with no analog in prior estimation paradigms — one shaped by non-determinism, context accumulation, and variable human oversight rather than by raw coding throughput. Existing cost estimation models — COCOMO II, Function Points, Use Case Points, Story Points — were built for a paradigm where this shift had not yet occurred, and none of them offers a mechanism for translating a project specification into a forecast of token consumption or the redirected human effort now spent on architecture, oversight, and correction.

This paper puts forward ACEM as an early-stage proposal for the software engineering community: a structured cost estimation framework that decomposes agentic development cost into LLM, HITL, and infrastructure components, and contributes three new constructs — the Revision Factor, the Context Factor, and the HITL Intensity Score — purpose-built to capture the dynamics that distinguish agentic pipelines from traditional development. Critically, ACEM treats LLM cost and human cost as two distinct, additive dimensions rather than as a simple substitution of one for the other: token consumption is an added cost the agentic paradigm introduces, while the HITL component reflects human effort that has been reshaped rather than removed — shifted upstream into architecture and requirements framing, and downstream into review, validation, and correction of agent output. By further establishing formal mappings from Use Case Points, Story Points, and Function Points to token-based cost drivers, ACEM allows organizations to build on their existing estimation practices rather than replace them outright.

We want to be explicit about what this paper does and does not claim. ACEM is offered here as a model structure and calibration methodology — not as a validated instrument. Its calibration constants are symbolic by design, and its predictive accuracy, generalizability, and practical usability remain open questions. Nor does this paper claim to know, in advance of calibration, how the balance between LLM cost and human cost will actually play out across task types and project scales — early illustrative estimates suggest this balance can shift substantially with task complexity and pipeline length, which is precisely the kind of question ACEM is designed to let the community answer empirically rather than assume. This is a deliberate first step rather than a finished result: we see greater value in placing a well-specified, falsifiable framework in front of the community early than in withholding it until a single study could validate it in isolation, which no individual research group is positioned to do given the diversity of agents, models, and domains in active use.

We therefore put ACEM forward as a proposal, and extend an open invitation to the community to validate, challenge, calibrate, and extend it — through the evaluation program outlined in Section 5 or through independent efforts of their own. Whether ACEM in its current form proves accurate, requires substantial revision, or is superseded by better alternatives, our hope is that it moves the field a step closer to a principled, shared basis for estimating cost in the agentic era of software engineering — one that accounts honestly for both the computational cost agents add and the human cost that agentic development redistributes rather than removes.